\documentclass[11pt,a4paper]{article}
\usepackage[latin1]{inputenc}
\usepackage[]{graphicx}
\usepackage[]{epsfig}

\newcommand{\vecomega}{\mbox{\boldmath$\omega$}}
\newcommand{\vecr}{\mbox{\boldmath$r$}}
\newcommand{\vece}{\mbox{\boldmath$e$}}
\newcommand{\vecf}{\mbox{\boldmath$f$}}
\newcommand{\vecv}{\mbox{\boldmath$v$}}

\begin{document}

\title{The physics of rotational flattening
and the point core model}
\author{Hanno Ess\'en\thanks{e-mail: hanno@mech.kth.se}
\\
Department of Mechanics, KTH \\    SE-100 44 Stockholm, Sweden}
\date{July 2002}

\maketitle

\begin{abstract}
A point mass at the center of an ellipsoidal homogeneous fluid is used
as a simple model to study the effect of rotation on the shape and
external gravitational field of planets and stars.  Maclaurin's
analytical result for a homogenous body is generalized to this model.
The absence of a third order term in the Taylor expansion of the
Maclaurin function leads to further simple but very accurate
analytical results connecting the three observables: oblateness
($\epsilon$), gravitational quadrupole ($J_{2}$), and angular velocity
parameter ($q$).  These are compared to observational data for the
planets.  The moments of inertia of the planets are calculated and
compared to published values.  The oblateness of the Sun is estimated.
Oscillations near equilibrium are studied within the model.  \\ \\
\end{abstract}

\newpage
\section{Introduction}
The rotation induced oblateness of astronomical bodies is a classical
problem in Newtonian and celestial mechanics (for the early history,
see Todhunter \cite{todhunter}).  It has twice played an important
role in the history of science.  In the early eighteenth century
measurements indicated a prolate shape of the Earth, in strong
conflict with the Newtonian prediction.  This was later shown to be
wrong by more careful measurements by Maupertuis, Clairaut, and
Celsius in northern Sweden in 1736.  Then, in 1967, measurements of
the solar oblateness were published and according to these it was much
larger than the Sun's surface angular velocity would explain.  The
confirmation of general relativity by Mercury's perihelion precession
would then be lost.  Also this problem is now gone and the modern
consensus is that the solar oblateness is too small to affect this
classic test of general relativity \cite{will,godier,laskar}.

The subject of the flattening of rotating astronomical bodies is thus
quite mature.  The classical theory is due mainly to Clairaut, Laplace,
and Lyapunov. Also Radau, Darwin, de Sitter and many others have made
important contributions. More recent accounts of the theory can be
found in, for example, Jeffreys \cite{jeffreys}, Zharkov et al.\
\cite{zharkov}, Cook \cite{cook}, Moritz \cite{moritz} and, partly,
in Chandrasekhar \cite{chandrasekhar}. Some pedagogical efforts
can be found in
Murray and Dermott \cite{murray}, or in Kaula \cite{kaula}.  As is
plain from these references the theory is quite involved.  Only the
unrealistic assumption that the body is homogeneous gives compact
analytical results.  Otherwise a specified radial density distribution
is needed and one resorts to cumbersome series (multipole) expansions,
or purely numerical methods, for quantitative results.  Here we will
present analytical results based on the assumption that the body
consists of a central point mass surrounded by a homogeneous fluid,
the so called point core model.  By varying the relative mass of the
fluid and the central point particle one can interpolate between the
extreme limits of Newton's homogeneous body and the Roche model
\cite{kippenhahn} with a dominating small heavy center.  The point
core model goes back to the work of G.~H.\ Darwin.  More recently it
has been used to study the shape of outer planet moons, see Hubbard
and Anderson \cite{hubbard}, Dermott and Thomas \cite{dermott}.

Apart from the basic point core approximation (i) several further
approximations are assumed here.  These are: (ii) that the shape is
determined by hydrostatic equilibrium and (iii) that the shape is
ellipsoidal.  These are not consistent. According to Hamy's theorem,
see Moritz \cite{moritz}, the exact shape is not ellipsoidal, so we
regard an ellipsoidal
shape as a constraint and find the equilibrium shape, among these, by
minimizing the energy.  A further approximation (iv) is the neglect of
differential rotation.  On the other hand fixed volume (``bulk
incompressibility'') need not be assumed; the equilibrium volume
problem separates from the shape problem.  In spite of these
approximations, which are standard in the literature, the mathematics
can be quite involved.  In this article I hope to clarify and simplify
it as much as possible.

\begin{figure}
\centering
\includegraphics[width=300pt]{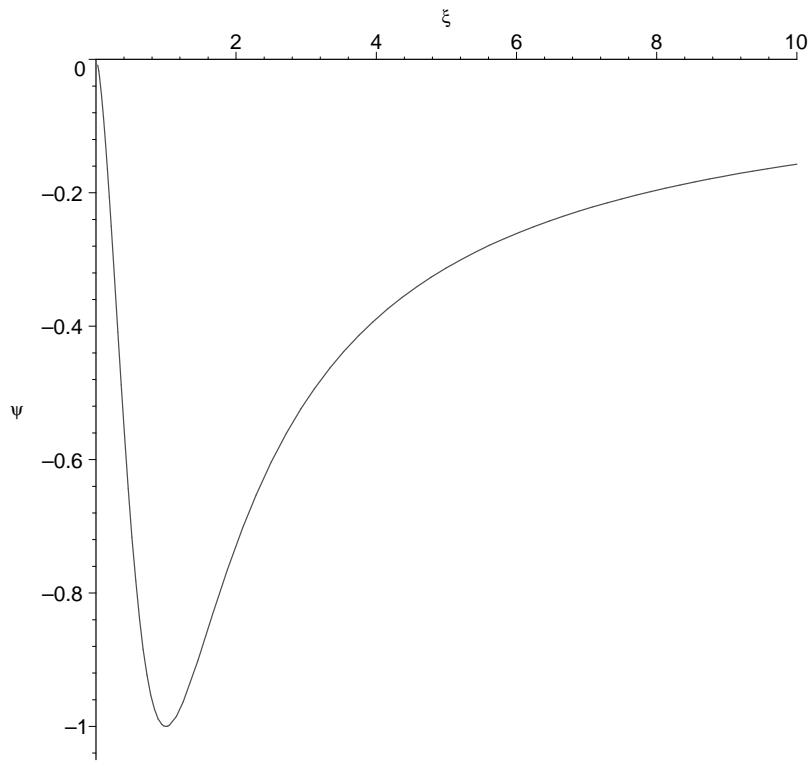}
\vspace{1ex}
\caption
{The function $\psi(\xi)$ of equation Eq.\ (\ref{eq.basic.phi.xi})
which gives the dimensionless gravitational energy of a homogeneous
spheroid with a central point mass. The minimum at $\xi=1$
corresponds to a sphere.
}
\label{FIG1}
\end{figure}

Mathematically our model then becomes a three degree of freedom
mechanical system for which we can calculate the kinetic and potential
(gravitational plus centrifugal) energies exactly.  Multipole
expansions in terms of spherical harmonics are not needed.  The three
degrees of freedom correspond to the three semi-axes of the ellipsoid
($a, b, c$) but these are transformed to three generalized coordinates
that describe size (or volume), $R$, spheroidal, $\xi$, and triaxial,
$\tau$, shape changes, see Eq.\ (\ref{eq.def.xi.eta}).  The statics
problem of equilibrium shape is solved by minimizing the potential
energy, $U(R,\xi,\tau)$ for a fixed $R$, Eq.\
(\ref{eq.energy.xi.tau}).  For slow rotation the shape will not be
triaxial so $\tau=0$.  Finding the shape, or flattening, is then only
a matter of minimizing a dimensionless potential energy,
\begin{equation}
\label{eq.dim.less.pot.en.xi}
u(\xi)=\psi(\xi)-\frac{1}{2}k \xi^2 ,
\end{equation}
given by the last two terms of Eq.\ (\ref{eq.energy.xi}).  Here
$\psi(\xi)$ is defined by Eq.\ (\ref{eq.basic.phi.xi}) and plotted in
Fig.\ \ref{FIG1} while $k$ is a constant that depends on rotational
parameter $q$ and dimensionless moment of inertia $\kappa^2$, Eq.\
(\ref{eq.kappa2.nu}).  The root $\xi(k)$ of the equation $\psi'(\xi)
-k\xi =0$ thus gives the flattening.  This root is to very high
accuracy given by
\begin{equation}
\label{eq.xi.root.of.k}
\xi(k)=\left[1-(7/2)\left(\sqrt{1+15 k/7} - 1\right) \right]^{-1/6} ,
\end{equation}
see Eqs.\ (\ref{eq.maclaurin.approx.inv}) and
(\ref{eq.eccentricity.relation.xi}), for the rotational parameters
that can be found in the solar system.

From this result, simple formulas, Eqs.\
(\ref{eq.maclaurin.approx.relation.q}) --
(\ref{eq.maclaurin.approx.relation.epsilon}), relating the
observables, rotation parameter, $q$, gravitational quad\-ru\-pole,
$J_{2}$, and excentricity squared, $e^2$, are obtained.  These appear
to be, partly, new, and their usefulness is demonstrated by comparing
with empirical data for the Sun and the rotating planets of the solar
system.  Variational methods have been used before to study similar
problems, see for example Abad et al.\ \cite{abad}. Denis et al.\
\cite{denis1} have pointed out that variational methods generally fail
to provide estimates of their accuracy. Therefore the agreement of our
formulas with empirical data, as demonstrated in Table /cite(table), is
important and demonstrates that our model catches the essential physics of
rotational flattening.

Finally small amplitude oscillations near the equilibrium are
investigated, starting from the Lagrangian
\begin{equation}
\label{eq.lagrangeian}
L=T(R,\xi,\tau,\dot R, \dot\xi,\dot\tau)-U(R,\xi,\tau),
\end{equation}
for our three degree of freedom model system.  This gives useful insight into
the physics of free stellar or planetary oscillations and their coupling to
rotation.  Our approximations are, however, too severe for these
results to be of quantitative interest.

\section{Basic geometric quantities}
\label{sec.ellipsoids}
Assume that $x, y, z$ are rectangular Cartesian coordinates in
three-dimen\-sion\-al space of a point with position vector
\begin{equation}
\label{eq.position.vect}
\vecr=x\vece_x + y \vece_y +z \vece_z .
\end{equation}
An ellipsoid, with semi-axes $a, b, c$, is the solid defined by
\begin{equation}
\label{eq.ellipsoid.Ec.def}
 \left[ \left(\frac{x}{a}\right)^2 +\left(\frac{y}{b}\right)^2
 +\left(\frac{z}{c}\right)^2 \right]^{1/2} \le 1 .
\end{equation}
If we put
\begin{equation}
\label{eq.def.xi.eta}
a=(\xi+\tau) R,\; b=(\xi-\tau) R,\; c=(\xi^2-\tau^2)^{-1} R
\end{equation}
and define
\begin{equation}
\label{eq.ellipsoid.E.def}
 E(\vecr;\xi, \tau) \equiv \left[ \left(\frac{x}{\xi+\tau}\right)^2
 +\left(\frac{y}{\xi-\tau}\right)^2
 +\left(\frac{z}{(\xi^2-\tau^2)^{-1}}\right)^2 \right]^{1/2}
\end{equation}
the ellipsoid is the set of points that fulfill
\begin{equation}
\label{eq.ellipsoid.def}
E(\vecr;\xi,\tau) \le R ,
\end{equation}
where $R$ is the geometric mean (or volumetric) radius
$R=\sqrt[3]{abc}$.  The formula for the volume of an ellipsoid now
gives
\begin{equation}
\label{eq.volume.notation}
V ={4\pi abc}/{3}={4\pi R^3}/{3},
\end{equation}
and we see that changes of $\xi$ and $\tau$ do not change the volume,
only the shape of the ellipsoid.

In what follows we will also consider the special case of spheroids.
A spheroid is an ellipsoid with two equal semi-axes.  We take $a=b\neq
c$.  This means that we take $\tau=0$ in the formulas above so that
\begin{equation}
\label{eq.def.xi.spheroid}
a=b=\xi R ,\; c=\xi^{-2} R,
\end{equation}
and (\ref{eq.ellipsoid.E.def}) becomes
\begin{equation}
\label{eq.spheroid.S.def}
E(\vecr;\xi,0)= \left[
\left(\frac{x}{\xi}\right)^2 +\left(\frac{y}{\xi}\right)^2
+\left(\frac{z}{\xi^{-2}}\right)^2 \right]^{1/2}.
\end{equation}
A spheroid is thus defined as the set of points $\vecr$ that fulfill
\begin{equation}
\label{eq.spheroid.def}
E(\vecr;\xi,0) \le R .
\end{equation}
Here $R$ is the (geometric) mean radius and $\xi$ is a parameter that
determines the shape in such a way that $\xi<1$ corresponds to a
prolate spheroid, $\xi=1$ to a sphere, and $\xi>1$ to an oblate, or
flattened, spheroid.

More familiar parameters used to define the shape of a spheroid are
the ellipticity $\epsilon$ and the eccentricity $e$.  The ellipticity
is defined by
\begin{equation}
\label{eq.ellipticity.def}
\epsilon \equiv \frac{a-c}{a} = 1 -\xi^{-3} .
\end{equation}
This is sometimes also called the (geometric) oblateness or the
flattening (denoted $f$).  Solving this equation, and expanding around
$\epsilon=0$, we have
\begin{equation}
\label{eq.ellipticity.relation.xi}
\xi = \frac{1}{\sqrt[3]{1-\epsilon}}=1+\frac{1}{3}\epsilon
+\frac{2}{9}\epsilon^2+O(\epsilon^3).
\end{equation}
One notes that $\epsilon$ is positive for oblate and negative for
prolate spheroids respectively.  The eccentricity, $e\ge 0$, is
defined as the usual eccentricity of the ellipse that is the
intersection of the spheroid and a plane containing the z-axis.  One
finds that
\begin{equation}
\label{eq.eccentricity.relation.xi}
 e^2 \equiv \frac{a^2-c^2}{a^2}= 1-\xi^{- 6}
\end{equation}
gives the eccentricity of these ellipses in the oblate case.  For the
prolate case $e^2 = 1-\xi^{ 6}$.

\section{The mass distribution}
Consider a non-rotating spherically symmetric body consisting of point
particles with masses $m_i$ and position vectors
\begin{equation}
\label{eq.position.vectors}
\vecr^0_i = x^0_i \vece_x + y^0_i\vece_y + z^0_i \vece_z .
\end{equation}
Assume that rotation induces a deviation of the positions so that the
new positions, $\vecr_i = \vecf(\vecr^0_i)$, are given by
\begin{equation}
\label{eq.position.vectors.dep}
\vecr_i(\xi,\tau)=x_i \vece_x + y_i \vece_y +z_i \vece_z=
(\xi+\tau) x^0_i \vece_x +(\xi-\tau) y^0_i\vece_y
+(\xi^2-\tau^2)^{-1} z^0_i \vece_z .
\end{equation}
This implies that we assume that the elastic displacement field can be
parameterized by the shape parameters $\xi, \tau$ with non-rotating
positions corresponding to $\xi=1, \tau=0$.  One notes that points
$\vecr^0$ that obeyed $E(\vecr^0; 1,0)= r,$ i.e.\ were found on a
sphere of radius $r\le R$, move to the surface given by
\begin{equation}
\label{eq.deformed.sphere}
E(\vecr;\xi,\tau) = r .
\end{equation}
The sphere is thus assumed to deform to an ellipsoid.

Assume that a body initially is spherically symmetric and has a mass
density $\varrho_{0}(r)$, where $r$ is the ordinary distance from the
center.  It is tempting to assume that a natural deformation of the
body when it starts to rotate is to a spheroidal shape in such a way
that the originally spherical equidensity surfaces,
$\varrho_{0}(r)=\mbox{constant}$, deform to similar spheroidal
surfaces given by $E(\vecr;\xi,0)=r$.  Unfortunately this is an
approximation for real bodies.  For a density that increases towards
the center, the ellipticity of the equidensity surfaces decrease with
$r$, in a manner described by Clairaut's equation, see
\cite{jeffreys,zharkov,cook,denis}.
Note, however, that for a constant density, or a constant density with
a central point mass, the assumption is not approximate, since any
surface is an equidensity surface, only the shape of the outer surface
matters.

Thus we now assume such a mass distribution.  To be precise we assume
that a mass $M_1$ is distributed with constant density
\begin{equation}
\label{eq.rho.h}
\varrho_1 \equiv  M_1/V
\end{equation}
inside the surface, $E(\vecr;\xi,\tau)=R$, and that the point mass
$M_0$ is at the center, $\vecr={\bf 0}$, of the ellipsoid.  The
density $\varrho$ is thus given by
\begin{equation}
\label{eq.mass.distribution}
\varrho(\vecr)= \left\{ \begin{array}{ll}
M_0 \delta_3(\vecr) + \varrho_1 \hskip 0.5cm & \mbox{if $E(\vecr;\xi,\tau)\le R$} \\
0 & \mbox{otherwise}
\end{array} \right.
\end{equation}
where $\delta_3$ is the three-dimensional Dirac delta function.

\section{Moments of inertia and quadrupole tensor}
If we put $m_i$ for the masses of the particles, the inertia tensor of
the body is given by $ I_{\alpha\beta} = \sum_i m_i (x_{\gamma i}
x_{\gamma i} \,\delta_{\alpha\beta}-x_{\alpha i}x_{\beta i}), $ where
we have put, $x_i =x_{1 i},\, y_i =x_{2 i},\, z_i =x_{3 i}$, and
$\delta_{\alpha\beta}$ is the usual Kronecker delta.  The inertia
tensor for the non-rotating body is diagonal with all diagonal
elements equal and given by
\begin{equation}
\label{eq.def.mom.of.inert}
I \equiv M(\kappa R)^2 =I^0_{z z}=\sum_i m_i [(x^0_i)^2 + (y^0_i)^2].
\end{equation}
Here we have introduced the total mass,
\begin{equation}
\label{eq.tot.mass.def}
M\equiv \sum_i m_i =M_0 + M_1 ,
\end{equation}
as well as the dimensionless radius of gyration $\kappa$.  The square
of this parameter, $\kappa^2$, also appears in the literature denoted
$I/MR^2$ (or $C/Ma^2$, or simply $k$) and is also called the moment of
inertia coefficient.

We calculate the dimensionless moment of inertia $\kappa^2$ for the
undeformed $\varrho$ of Eq.\ (\ref{eq.mass.distribution}).  Using
Eq.\ (\ref{eq.def.mom.of.inert}) and $r^2=x^2+y^2+z^2$ we find
\begin{equation}
\label{eq.mom.of.inertia}
I=M(\kappa R)^2= \int_{r\le R}\!\!  \varrho_{1}(x^2+y^2) {\rm d}V =
\frac{2}{3} \int_0^{R} \frac{3 M_1}{4\pi R^3} r^2 4\pi r^2 {\rm d} r
= \frac{2}{5} M_1 R^2,
\end{equation}
where ${\rm d}V=4\pi r^2 {\rm d}r$ and Eq.\ (\ref{eq.rho.h}) have
been used.  This is the usual moment of inertia for a solid sphere of
mass $M_{1}$.  This implies that
\begin{equation}
\label{eq.kappa2.of.Ms}
\kappa^2=\frac{2}{5}\frac{M_1}{M} = \frac{2}{5}\frac{M-M_{0}}{M}
\end{equation}
is the squared (dimensionless) radius of gyration of the undeformed
body.  Note that $0\le \kappa^2 \le 2/5$ since the model can go
between the limits of a point ($\kappa^2 \rightarrow 0$) and a
homogeneous sphere ($\kappa^2=2/5=0.4$).

For the deformed body one gets, using (\ref{eq.position.vectors.dep}),
\begin{equation}
\label{eq.mom.of.inert.kappa}
I_{zz}=\sum_i m_i ({x}^2_i + {y}^2_i) = M(\kappa R)^2 (\xi^2 +\tau^2)
= I (\xi^2 +\tau^2) .
\end{equation}
$I_{zz}$ is often denoted by $C$.  For spheroids ($\tau=0$) the other
moments of inertia become $ I_{xx} =I_{yy}=\frac{1}{2}I
(1+\xi^{-6})\xi^2 .  $ For these one frequently finds the notation
$A$, and $B$, respectively, in the literature.  The quadrupole tensor,
$D_{\alpha\beta}$, is given in terms of the inertia tensor by $
D_{\alpha\beta}=I_{\gamma\gamma}\, \delta_{\alpha\beta} -3
I_{\alpha\beta}.  $ It is identically zero for the undeformed body and
diagonal for the same axes as the inertia tensor.  For the spheroidal
($\tau=0$) body its components are given by
\begin{equation}
\label{eq.quadrupole.comp.xx}
D \equiv D_{xx} = D_{yy} = \frac{1}{2} I (1-\xi^{-6}) \xi^2 ,
\end{equation}
and $D_{zz} =-2 D.$ Thus $D=I_{zz}-I_{xx}$ (or $D =C-A)$ is positive
for oblate and negative for prolate shapes.

\section{Potential and hydrostatic equilibrium}
One measure of the oblate shape that is of great interest is the, so
called, gravitational quadrupole $J_2$.  This dimensionless quantity,
gives the deviation of the gravitational potential, $\phi$, outside
the body, from the simple $1/r$-dependence of spherically symmetric
bodies.  The first two terms in the multipole expansion of this
potential are, see e.g.\ Stacey \cite{stacey},
\begin{equation}
\label{eq.grav.pot.J2}
\phi(r,\vartheta) =-\frac{GM}{r}\left[1-J_2\left( \frac{a}{r}
\right)^2 \frac{3\cos^2\vartheta -1}{2} \right],
\end{equation}
where $\vartheta$ is the polar angle (colatitude) in spherical
coordinates, $M$ the total mass, and $a$ the equatorial radius of the
body.  One can show \cite{stacey} that $J_2$ of formula
(\ref{eq.grav.pot.J2}) is given by $J_2= D/(Ma^2)$.  Use of $D$ given
in (\ref{eq.quadrupole.comp.xx}), and $a=\xi R$, then gives
\begin{equation}
\label{eq.J2.expression}
J_2 \equiv \frac{D}{Ma^2} =\frac{1}{2}(1-\xi^{-6}) \kappa^2 =
\frac{1}{2} \kappa^2 e^2 ,
\end{equation}
for the quadrupole.  Here we have also used $a=\xi R$ and Eq.\
(\ref{eq.eccentricity.relation.xi}).  Note that $J_2=J_2(e^2)$ is
independent of $R$, but directly proportional to the eccentricity
squared.

Assuming the oblateness to be small, and due to hydrostatic
equilibrium in the combined gravitational and centrifugal force
fields, one can derive a formula connecting $J_2$ with the angular
velocity and the ellipticity.  One uses that the shape of the surface
is determined by the hydrostatic equilibrium equation
\begin{equation}
\label{eq.hydrostat.equilibr}
\nabla p + \varrho \nabla (\phi +\phi_c)=0,
\end{equation}
where $p$ is pressure, $\varrho$ mass density, and
\begin{equation}
\label{eq.grav.pot.in.rot.syst}
\phi_c(r,\vartheta) =-\frac{1}{2}\omega^2 r^2\sin^2\vartheta
\end{equation}
is the potential of the centrifugal force in a system rotating with
angular velocity $\omega$ about the z-axis.  This means that the
surface of the body must be an equipotential surface of $\phi+\phi_c$.
The constancy of $\phi(r,\vartheta)+\phi_c(r,\vartheta)$ on the
surface of the body can be used to derive the approximate relationship
\begin{equation}
\label{eq.J2.alpha.epsilon}
\epsilon = \frac{3}{2} J_2 + \frac{1}{2} q
\end{equation}
\cite{cook}.  Here we have introduced the ratio $q$ of equatorial
centrifugal to gravitational acceleration,
\begin{equation}
\label{eq.acc.centrifug.grav.ratio}
q \equiv \frac{\omega^2 R}{g}=\frac{\omega^2 R}{GM/R^2}
=\frac{R^3\omega^2}{GM}=\frac{3\omega^2}{4\pi G \bar{\varrho}}.
\end{equation}
Here $\bar{\varrho}=M/V$ is the mean density.  Instead of $q$ the
notation $m$ occurs in the literature, see Zharkov et al.\
\cite{zharkov}, who reserve $q$ for the corresponding quantity
with the equatorial radius $a$ replacing the mean radius $R$ (recall
$R^3=a^2 c$).  The advantage of the definition
(\ref{eq.acc.centrifug.grav.ratio}) is that it makes $q$ independent
of flattening.  The relationship (\ref{eq.J2.alpha.epsilon}), is only
valid to first order in small quantities, so here the difference does
not matter, but we will go to higher order below, so take notice.
This relationship between the three quantities $q$ (or $m$),
$\epsilon$ (or $f$), and $J_{2}$, can thus be used to test the
hypothesis of hydrostatic equilibrium empirically.  This will be done
below, using more exact results.  For Jupiter and Saturn these will
improve significantly on (\ref{eq.J2.alpha.epsilon}).

\section{Rotational and gravitational energy}
To study a rotating body, we go to the rotating system, and assume
that all particles are affected by the centrifugal and gravitational
potential energies.  If the system is rotating rigidly about the
z-axis with angular velocity vector $ \vecomega = \omega\,
\vece_z, $ the potential energy of particle $i$ in the centrifugal
force field is $m_i \phi_c(r_i,\vartheta_i)$.  The total centrifugal
potential energy of the body (system of particles) is
\begin{equation}
\label{eq.centrifug.energy.0}
\Phi_c = -\frac{1}{2}\sum_{i} m_i(x_i^2+y_i^2) \omega^2 = -\frac{1}{2}
I_{zz} \omega^2.
\end{equation}
This is the negative of the rotational kinetic energy.  Using
(\ref{eq.mom.of.inert.kappa}) we can write
\begin{equation}
\label{eq.centrifug.energy.1}
\Phi_c =-\frac{1}{2} I (\xi^2+\tau^2) \omega^2 = - \frac{1}{2}
M(\kappa R)^2 (\xi^2+\tau^2) \omega^2
\label{eq.centrifug.energy}
\end{equation}
for the total potential energy of the particles in the centrifugal
potential.

The gravitational potential from an ellipsoid
(\ref{eq.ellipsoid.Ec.def}), of constant density $\varrho_1$, at an
interior point is  \cite{macmillan}
\begin{equation}
\label{eq.grav.pot.of.hom.ellipsoid.interior}
\phi(x,y,z) =-G\varrho_1 \pi abc \int_{0}^{\infty} \left(
1-\frac{x^2}{a^2+s}-\frac{y^2}{b^2+s}-\frac{z^2}{c^2+s} \right)
\frac{{\rm d}s}{P(s)},
\end{equation}
where $P(s)=\sqrt{(a^2+s)(b^2+s)(c^2+s)}$.  Since $M_0 \phi(0,0,0)$ is
the interaction energy of the homogeneous ellipsoid with a point
particle of mass $M_0$ at the origin one finds the expression
\begin{equation}
\label{eq.grav.pot.energy.of.hom.ellipsoid.center.particle} \Phi_0
=-GM_0 M_1 \frac{3}{4} \int_{0}^{\infty} \frac{{\rm d}s}{P(s)}
\end{equation}
for this interaction energy.  Here $abc=R^3$ and Eq.\ (\ref{eq.rho.h}) has been used.

An elementary calculation based on
(\ref{eq.grav.pot.of.hom.ellipsoid.interior}), see Landau and Lifshitz
\cite{ll2}, also shows that the gravitational self energy of the
ellipsoid is
\begin{equation}
\label{eq.grav.pot.energy.of.hom.ellipsoid.self}
\Phi_1 =-GM_1^2 \frac{3}{10} \int_{0}^{\infty} \frac{{\rm d}s}{P(s)}.
\end{equation}
Use of the definitions (\ref{eq.def.xi.eta}) and the substitution
$s=R^2 u$ gives
\begin{equation}
\label{eq.grav.pot.energy.ellipsoid.integral} \int_{0}^{\infty}
\frac{{\rm d}s}{P(s)} =\frac{1}{R} \int_{0}^{\infty} \frac{{\rm d}u}{
\sqrt{[(\xi+\tau)^2+u][(\xi-\tau)^2+u][(\xi^{2}-\tau^{2})^{-2}+u]}}.
\end{equation}
If we define
\begin{equation}
\label{eq.grav.energy.triaxial.chi}
\chi(\xi,\tau)=-\frac{1}{2}\int_{0}^{\infty} \frac{{\rm d}s }{
\sqrt{[(\xi+\tau)^2+s][(\xi-\tau)^2+s][(\xi^{2}-\tau^{2})^{-2}+s]}},
\end{equation}
we find that the total gravitational energy of the mass distribution
(\ref{eq.mass.distribution}) is given by
\begin{equation}
\label{eq.tot.grav.energ}
\Phi =\Phi_1 + \Phi_0 =\frac{GM_1}{R} \left( \frac{3 M_1}{5} +\frac{3
M_0}{2} \right)\chi(\xi,\tau),
\end{equation}
where $\chi(1,0)=-1$.  We define the constant $\nu$, the dimensionless
gravitational radius, through
\begin{equation}
\label{eq.grav.energy.triaxial.2}
 \Phi=\frac{GM^2}{R \nu} \chi(\xi,\tau).
\end{equation}
Comparing with (\ref{eq.tot.grav.energ}) and using $M=M_0+M_1$
together with the result (\ref{eq.kappa2.of.Ms}), a small calculation
reveals that the dimensionless gravitational radius, $\nu$, is given
by
\begin{equation}
\label{eq.tot.grav.energ.nu.ch} {1}/{\nu}
=({15}/{8})\, \kappa^2 (2-3\kappa^2).
\end{equation}
It is interesting to note that the gravitational energy of a density
(\ref{eq.mass.distribution}), with fixed mass $M$ and radius $R$, is
minimized for $\kappa^2=1/3$ (this is incidentally very close Earth's
value).

Using the definition (\ref{eq.grav.energy.triaxial.chi}) it is easy to
find the Taylor expansion of $\chi$ around $\xi=1, \tau=0$.  This
gives
\begin{eqnarray}
\label{eq.grav.energy.triaxial.chi.series}
\chi(\xi,\tau)= -1+\frac{4}{5}(\xi-1)^2+\frac{4}{15}\tau^2+\ldots
\end{eqnarray}
so the sphere really corresponds to a minimum in the gravitational
energy.  Higher order terms are easily generated by computer algebra
programs, but we will not need them here.  To study the behavior
further away from the minimum one might use, instead of
(\ref{eq.grav.energy.triaxial.chi}), an excellent Pad\'e approximation
of the ellipsoidal energy derived by Dankova and Rosensteel
\cite{dankova}.  Near the spherical minimum it is almost
indistinguishable from the exact expression.

For spheroids we define
\begin{equation}
\label{eq.def.psi.chi}
\psi(\xi) =\chi(\xi,0)
\end{equation}
to be the dimensionless gravitational potential energy.  For $\tau=0$
the integral in (\ref{eq.grav.energy.triaxial.chi}) can be found in
terms of elementary functions.  Several different alternative
expressions exist and they are always different depending on whether
$\xi$ is greater or less than unity.  However, noting that, ${\rm
arctanh}\; {\rm i}z ={\rm i} \arctan z$, the two expressions in
\begin{equation}
\label{eq.basic.phi.xi}
\psi(\xi) = \left\{ \begin{array}{ll} \displaystyle - \frac{\xi^2 {\rm
arctanh} \sqrt{1-\xi^6}}{\sqrt{1-\xi^6}} & \mbox{ for $0< \xi\le 1\;$
(prolate case)} \\
& \\
\displaystyle - \frac{\xi^2 \arctan \sqrt{\xi^6-1}}{\sqrt{\xi^6-1}} &
\mbox{ for $1\le \xi <\infty\;$ (oblate case).}
\end{array}
\right.
\end{equation}
are equivalent and can be used for all $0 \le \xi \le \infty$.  These
are really the same real function if only the correct branch is chosen
when passing through $\xi=1$.  Series expansion, of either
alternative, around $\xi=1$ gives
\begin{equation}
\label{eq.sereis.ph}
\psi(\xi) = -1 +\frac{4}{5}(\xi -1)^2 -\frac{92}{105}(\xi-1)^3
+\frac{29}{105}(\xi-1)^4+O[(\xi-1)^5].
\end{equation}
This function, which is shown in Figure \ref{FIG1}, has the
following properties.  The prolate limit, $\xi\rightarrow 0$, of an
infinite line is zero: $\lim_{\xi\rightarrow 0} \psi(\xi) =0$.  The
oblate limit, $\xi\rightarrow \infty$, an infinite circular disc, is
also zero: $\lim_{\xi\rightarrow \infty} \psi(\xi) =0$.  The sphere
($\xi=1$) corresponds to a minimum of $\psi(\xi)$ so that
$\psi(1)=-1$, $\psi'(1)=0$, and $\psi''(1)=8/5$.

\section{Minimizing the energy}
The total energy, $U$, of a static body (star or planet) will, apart
from the centrifugal energy, $\Phi_c$ of Eq.\ (\ref{eq.centrifug.energy.1}), and gravitational energy, $\Phi$ of
Eq.\ (\ref{eq.grav.energy.triaxial.2}), also consist of some
volume dependent energy, $\Phi_{v}(R)$.  We can then write this total
energy it in the form
\begin{equation}
\label{eq.energy.xi.tau}
 U(R,\xi,\tau)= \Phi_{v}(R)+\frac{GM^2}{R \nu}\chi(\xi,\tau)
 -\frac{1}{2} M(\kappa R)^2 \omega^2 (\xi^2 +\tau^2) .
\end{equation}
By minimizing this energy we can find out whether a system tends to be
triaxial or spheroidal.  For small angular velocities the spheroidal
case is known to be relevant \cite{chandrasekhar,dankova}.  We thus put
$\tau=0$ here and  get
\begin{equation}
\label{eq.energy.xi}
 U(R,\xi)= \Phi_{v}(R)+\frac{GM^2}{R \nu}\psi(\xi) -\frac{1}{2}
 M(\kappa R)^2 \omega^2 \xi^2 .
\end{equation}
Using this we can find out if the body will be prolate or oblate and
how this is affected by rotation.  Since we concentrate on stars and
planets surface forces and other neglected shape dependent forces are
normally very small compared to the gravitational and centrifugal
energies.

The equilibrium equations can be written
\begin{equation}
\label{eq.dE.dl}
 \frac{\partial U}{\partial R}= \frac{ {\rm d} \Phi_{v}}{{\rm d} R}
 -\frac{GM^2}{R^2\nu}\psi(\xi) -M \kappa^2 R \omega^2 \xi^2 =0 ,
\end{equation}
\begin{equation}
\label{eq.dE.dxi}
\frac{\partial U}{\partial \xi}= \frac{GM^2}{R\nu} \psi'(\xi)-M
(\kappa R)^2\omega^2 \xi =0,
\end{equation}
where we have denoted differentiation by $\xi$ with a prime.  If we
multiply the first equation by $R$ and the second by $\xi$ and
subtract we get, after rearrangement,
\begin{equation}
\label{eq.dilatation.formula}
\frac{{\rm d} \Phi_{v}}{{\rm d}R}= \frac{GM^2}{R^2\nu}\left[\psi(\xi)
+\psi'(\xi)\xi\right].
\end{equation}
This is essentially an equation of hydrostatic equilibrium.  Note that
it determines the equilibrium $R$-value with no direct reference to
the angular velocity $\omega$.  The second equation,
(\ref{eq.dE.dxi}), can be written
\begin{equation}
\label{eq.force.bal.grav.surf.centrifug}
\frac{GM^2}{R\nu}\psi'(\xi)=M(\kappa R)^2\omega^2 \xi,
\end{equation}
and this expresses a balance of the shape dependent part of the
gravitational force with the centrifugal force.  If we put
\begin{equation}
\label{eq.def.k}
k \equiv \frac{R^3\omega^2}{GM} \nu\kappa^2 =q\nu\kappa^2 ,
\end{equation}
this turns into the concise expression
\begin{equation}
\label{eq.force.bal.grav.surf.centrifug.2}
\psi'(\xi) = k \xi.
\end{equation}
Since $\psi'(1)=0$ this will cause a shift of the shape from the
spherical equilibrium, at $\xi=1$, to $\xi>1$, as long as $\omega >0$.
Use of (\ref{eq.tot.grav.energ.nu.ch}) shows that
\begin{equation}
\label{eq.kappa2.nu}
k=({8}/{15})\, {q}/({2-3\kappa^2})
\end{equation}
The equilibrium $\xi$ is thus entirely determined by $q$ and the
moment of inertia $\kappa^2$.

\section{Finding the position of the minimum}
To investigate Eq.\ (\ref {eq.force.bal.grav.surf.centrifug.2}) we
use the second variant for $\psi(\xi)$ in Eq.\ (\ref{eq.basic.phi.xi}) and get, after some algebra,
\begin{equation}
\label{eq.force.bal.grav.surf.centrifug.3}
k =(\xi^6-1)^{-\frac{3}{2}} \left[ (3+\xi^6-1)
\arctan(\sqrt{\xi^6-1})-3\sqrt{\xi^6-1} \right] .
\end{equation}
This looks simpler if we introduce the new variable
$\eta(\xi)=\sqrt{\xi^6-1}$.  Use of Eq.\ (\ref{eq.eccentricity.relation.xi}) shows that
\begin{equation}
\label{eq.def.of.eta.eps.e}
\eta \equiv \sqrt{\xi^6-1} = e/\sqrt{1-e^2},
\end{equation}
so that $\eta$ can also be regarded as a function of the eccentricity,
$\eta(e)$.  Now we get
\begin{equation}
\label{eq.maclaurin}
k= \left[ (3+\eta^2) \arctan\eta-3\eta \right]/\eta^3 .
\end{equation}
For the homogeneous case $k =2q/3$ and this formula, written in terms
of the eccentricity by means of the identity,
$\arctan(e/\sqrt{1-e^2})=\arcsin e$, is equivalent to the celebrated
Maclaurin's formula \cite{chandrasekhar} from 1742.

\begin{figure}
\centering
\includegraphics[width=300pt]{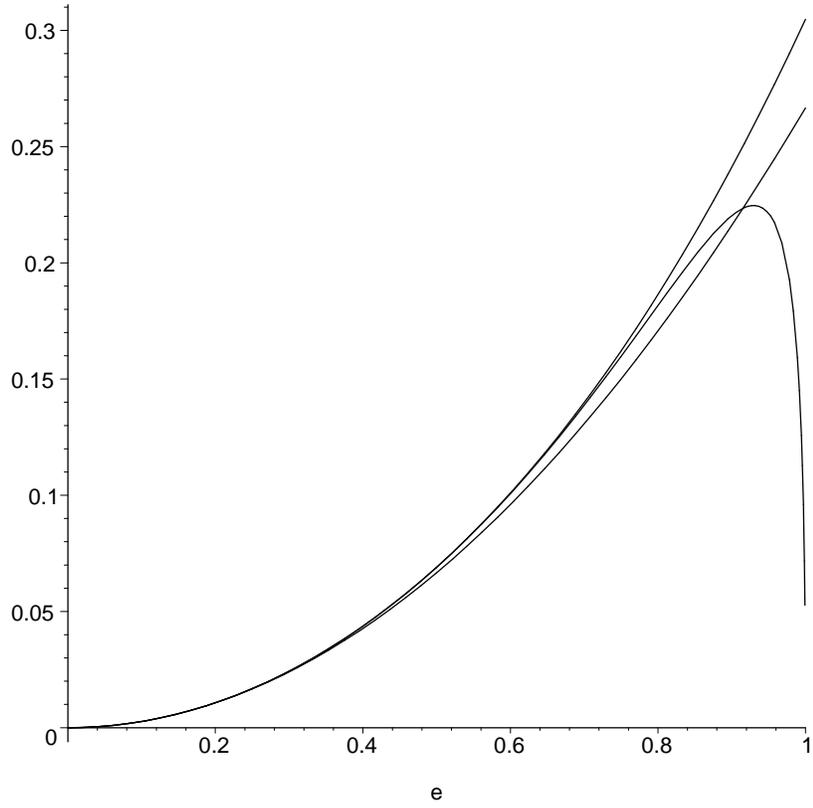}
\vspace{1ex}
\caption
{The Maclaurin function $f(e)=[(3+\eta^2)\arctan\eta-3\eta]/\eta^3$
with $\eta=e/\sqrt{1-e^2}$ compared to the approximations $4e^2/15$
(lower curve) and $(4e^2/15)(1+e^2/7)$ (upper curve).
}
\label{FIG2}
\end{figure}

Taylor expansion of the Maclaurin function around $e=0$ only contains
positive even powers of $e$.  The coefficient of $e^6$ vanishes so a
very good approximation for this function is implied by
\begin{equation}
\label{eq.maclaurin.approx}
\left[ (3+\eta^2) \arctan\eta-3\eta \right]/\eta^3 = \frac{4}{15} e^2
\left( 1+\frac{1}{7}e^2 \right) + O(e^8).
\end{equation}
The two term approximation is nearly perfect for all planetary
$e$-values, see Figure \ref{FIG2}.  Saturn has maximum $e=0.43$ and the absolute error at
this value is $2\cdot 10^{-5}$ and the relative error is $4\cdot
10^{-4}$.  The relative error even remains below $1\%$ up to $e=0.75$.
Using this approximation we must solve
\begin{equation}
\label{eq.maclaurin.approx.eq}
k = \frac{4}{15} e^2 ( 1+\frac{1}{7}e^2 )
\end{equation}
 and solving this for $e^2$ gives
\begin{equation}
\label{eq.maclaurin.approx.inv}
e^2= \frac{7}{2} \left( \sqrt{1+\frac{15}{7}k }-1\right).
\end{equation}
Use of Eq.\ (\ref{eq.J2.expression}) gives $\kappa^2=2J_{2}/e^2$.
Insertion of this into (\ref{eq.kappa2.nu}) eliminates $\kappa^2$ from
$k=\nu\kappa^2 q$.  Some algebra then leads to the equation
\begin{equation}
\label{eq.maclaurin.approx.relation.q}
q(e^2,J_{2})=\left(1+ e^2/7 \right) \left(e^2- 3J_2\right)
\end{equation}
for $q$.  This equation can be solved for $J_{2}$ or for $e^2$ to give
\begin{eqnarray}
\label{eq.maclaurin.approx.relation.J2}
J_{2}(q,e^2)&=&\frac{1}{3}\left( e^2-\frac{q}{1+e^2/7}
\right), \\
\label{eq.maclaurin.approx.relation.epsilon}
e^2(q,J_{2})&=&\frac{1}{2}\left[ 3 J_2+ 7 \left( \sqrt{
\left( 1+\frac{3}{7} J_2\right)^2 +\frac{4}{7} q }-1 \right)
\right],
\end{eqnarray}
respectively.  These results, together with
\begin{eqnarray}
\label{eq.e2.eps}
e^2&=&\epsilon(2-\epsilon) \approx 2\epsilon, \\
\label{eq.eps.e2}
\epsilon&=&1-\sqrt{1-e^2} ,
\end{eqnarray}
are thus more exact versions of Eq.\ (\ref{eq.J2.alpha.epsilon}).
Already the approximation $4e^2/15$, to the Maclaurin function, gives
only a $3\%$ error for Saturn, and simpler formulae, but use of these
more exact versions removes all numerical considerations from the
problem.

\section{Theory versus observational data}
Table~\ref{table} compares modern data with the predictions of Eqs.\
(\ref{eq.maclaurin.approx.relation.q}) -- (\ref{eq.eps.e2}).  Observed
data for $q$, $\epsilon$, and $J_{2}$ from Lodders and Fegley
\cite{lodders} are given in the first (horizontal) row of data for
each of the major rotating planets.  In the same line $e^2$ calculated
from $e^2=\epsilon(2-\epsilon)$ is then given.  The last value in the
row is the dimensionless moment of inertia $\kappa^2=I/MR^2$ as given
by Lodders and Fegley \cite{lodders}.  These values are based on
calculation.  For original sources the reader may consult The
Planetary Scientist's Companion by Lodders and Fegley \cite{lodders}.

For each pair of the values $q$, $J_{2}$, and $e^2$, in the first row,
the third value is calculated using
(\ref{eq.maclaurin.approx.relation.q}) --
(\ref{eq.maclaurin.approx.relation.epsilon}) and the results are
displayed in the second row of data for each planet in
Table~\ref{table}.  $\epsilon$ is then found from equation
(\ref{eq.eps.e2}). For Jupiter and Saturn there is a third row of data in
Table~\ref{table}.  This row is analogous to the second row except
that the data are calculated from observational pairs of data using
the first order formula (\ref{eq.J2.alpha.epsilon}) so that $q$ is
given by observational $\epsilon$ and $J_{2}$ according to
$q=2\epsilon-3J_{2}$, $\epsilon$ is found directly from
(\ref{eq.J2.alpha.epsilon}) using the observational $q$ and $J_{2}$
and so on.  These rows illustrate the fact that the non-linear
formulas of this work are more accurate than first order results, in
spite of the point core approximation.  Even for the Earth the first
order formula gives discrepancies in the third digit.

Table~\ref{table} shows clearly that the observational data and theory
are in reasonable agreement for the planets with the exception of Mars
and Uranus.  The perfect agreement for the Earth is probably due to
the fact that the oblateness is determined from the shape of the geoid
(the equipotential surface at sea-level).  For Mars the disagreement
may depend on the fact that oblateness is determined from geometrical
shape rather than from the shape of an equipotential surface.  Also,
of course, the assumption of hydrostatic equilibrium may not be
perfect for Mars.  The problems with the Uranus data, and to some
extent those for Neptune, are probably due to observational uncertainty.

\section{Oblateness and moment of inertia}
Instead of eliminating the moment of inertia (or mass ratio)
$\kappa^2$ of Eq.\ (\ref{eq.kappa2.of.Ms}) from the equations as done
above one can retain it and try to use observed data to get
information about $\kappa^2$.  Series expansion of Eq.\
(\ref{eq.maclaurin}) with $\eta$ expressed in terms of appropriate
variables gives
\begin{equation}
\label{eq.ellipticity.of.kappa2}
\epsilon =\frac{q}{2-3\kappa^2} + \frac{3}{14}\left(
\frac{q}{2-3\kappa^2} \right)^2+ O(q^3)
\end{equation}
for the flattening, or oblateness, expressed in terms of $q$ and the
dimensionless moment of inertia.  For a homogeneous body
$\kappa^2=2/5$ so $\epsilon=\frac{5}{4}q+\frac{75}{224}q^2+O(q^3)$ and
the old Newtonian result, $\epsilon=\frac{5}{4}q$, is obtained, to
first order.  In the opposite (Roche) limit of dominating central mass
($\kappa^2=0$) and one finds $\epsilon=\frac{1}{2}q
+\frac{3}{56}q^2+O(q^3)$.

In a similar way we get from $J_2$ of Eq.\ (\ref{eq.J2.expression})
that
\begin{equation}
\label{eq.J2.expression.3}
J_2 = \frac{\kappa^2 }{2-3\kappa^2} q+ \frac{8}{21}\left(
\frac{\kappa}{2-3\kappa^2} \right)^2q^2+ O(q^3)
\end{equation}
is the dimensionless quadrupole moment of the rotating body.  For the
homogeneous ($\kappa^2=2/5$) body this implies $J_2=\frac{1}{2}q
+\frac{8}{525}q^2 +O(q^3)$.  In the opposite limit of
$\kappa^2\rightarrow0$ one finds that $J_2$ goes to zero.

Our results (\ref{eq.ellipticity.of.kappa2}) and
(\ref{eq.J2.expression.3}), to first order, are
\begin{eqnarray}
\label{eq.ellipt.kappa2}
\epsilon(\kappa^2)=\frac{1}{2-3\kappa^2}q, \\
\label{eq.J2.kappa2}
J_2(\kappa^2) =\frac{\kappa^2}{2-3\kappa^2}q,
\end{eqnarray}
and imply that $ J_2=\epsilon\kappa^2$. These may be compared to
similar expressions derived by G.\ H.\ Darwin using the Radau
equation \cite{cook}
\begin{eqnarray}
\label{eq.ellipt.kappa2.darw}
\epsilon(\kappa^2)=\frac{40}{16+25\left(2-3\kappa^2\right)^2 } q, \\
\label{eq.J2.kappa2.darw}
J_{2}(\kappa^2) =\frac{64-25\left(2-3\kappa^2\right)^2 }{48+75
\left(2-3\kappa^2\right)^2 } q.
\end{eqnarray}
These two have the same values at $\kappa^2=2/5$, the homogeneous
sphere, and the same derivatives with respect to $\kappa^2$ at this
point, as (\ref{eq.ellipt.kappa2}) and (\ref{eq.J2.kappa2}),
respectively.  The simple expressions (\ref{eq.ellipt.kappa2}) and
(\ref{eq.J2.kappa2}), derived here, however, also have natural limits
for $\kappa^2=0$, in contrast to e.g.\ (\ref{eq.J2.kappa2.darw}), which
becomes negative for $\kappa^2<2/15$.

We now use Eq.\ (\ref{eq.J2.expression}) i.e.\ $\kappa^2=2J_{2}/e^2$
and Eq.\ (\ref{eq.maclaurin.approx.relation.epsilon}) for
$e^2(q,J_{2})$ to calculate $\kappa^2$ in terms of the observed $q$
and $J_{2}$.  The last entry in the second row for each planet of
Table~\ref{table} gives $\kappa^2$,
\begin{equation}
\label{eq.kappa2.of.J2.and.e2}
\kappa^2(q,J_{2})= 2 J_{2}/ e^2(q,J_{2}) ,
\end{equation}
as calculated from observational, first row, $q$ and $J_{2}$.  Since
$q$ and $J_{2}$ are easier to measure accurately than flattening this
should give more reliable values than using the observed ellipticity
directly.  In this way one obtains information about the radial mass
distribution in the interior of the planet from purely external data.
The resulting $\kappa^2$-values are given last in the second line for
each planet in Table~\ref{table}.  The data obtained in this way are
compared with published and tabulated $\kappa^2$ values for the
planets (Lodders and Fegley \cite{lodders}).  For Earth and Mars the
agreement is reassuring.  For the outer planets there is naturally a
lot of uncertainty but at least the increasing trend from a minimum at
Saturn is a common feature.

One problem with the outer planets, and with the Sun in particular, is
to decide what angular velocity, $\omega$, should be used to find $q$.
Since these bodies have differential rotation some average must be
used.  There is a well defined theoretical average angular velocity
(Ess\'en \cite{essen}) but it is not directly observable.  Once an
average angular velocity has been selected for the Sun one can use the
present theory and a theoretical $\kappa^2$-value ($0.059$) to
estimate the solar oblateness and $J_{2}$.  The definition of average
angular velocity in \cite{essen} indicates that an angular velocity
near the radius of gyration $\kappa R\approx 0.24 R$ should be
appropriate.  The further assumption that angular velocity is constant
on coaxial cylinder surfaces (Kippenhahn and Weigert
\cite{kippenhahn}, Ulrich and Hawkins \cite{ulrich}) indicates that an
angular velocity near the poles of the Sun rather than equatorial
angular velocity is relevant.

A handbook value \cite{lodders} for the polar angular velocity of the
Sun is $\omega=2.1\cdot 10^{-6}\,$rad s$^{-1}$.  Use of this gives the
$q$-value ($1.15\cdot 10^{-5}$) in Table~\ref{table}.  The precise
numbers are not important here.  What is important is that, because of
the small $\kappa^2$, an angular velocity near the pole rather than an
equatorial angular velocity is used.  Otherwise the flattening will be
exaggerated.  Use of $q=1.15\cdot 10^{-5}$ and $\kappa^2=0.059$ from
Lodders and Fegley \cite{lodders}, together with formulas
(\ref{eq.ellipt.kappa2}) and (\ref{eq.J2.kappa2}), gives the results
in the top row of Table~\ref{table}, that is, $\epsilon \sim 6.3\cdot
10^{-6}$, and $J_{2} \sim 3.7\cdot 10^{-7}$.  These values are, at
least, of the same order of magnitude as the currently best motivated
values \cite{godier,laskar,ulrich,paterno,lydon} which are roughly
$\epsilon \sim 9 \cdot 10^{-6}$ and $J_{2} \sim 2\cdot 10^{-7}$.

\section{Small oscillations near equilibrium}
We now wish to study the small amplitude motions of a gravitating
rotating nearly spherical body.  We wish to know how the radius $R$ is
affected by the rotation so we take $R$ to denote the constant
non-rotating equilibrium radius and let $R\lambda$ be the variable
radius.  $\lambda$ is thus a dimensionless variable which is equal to
unity at the undeformed geometry.  We parameterize the positions of
the particles of the system as follows
\begin{equation}
\label{eq.position.vectors.dep.2}
\vecr_i(\lambda, \xi, \tau)=R \lambda [ (\xi+\tau) x^*_i \vece_x
+(\xi-\tau) y^*_i\vece_y +(\xi^2- \tau^2)^{-1} z^*_i \vece_z] .
\end{equation}
The $x^*_i, y^*_i, z^*_i$ are then the non-rotating equilibrium
dimensionless position coordinates of the particles.  Equilibrium
corresponds to $\lambda=\xi=1$, $\tau=0$, see Section
\ref{sec.ellipsoids}.  $\lambda$ is a dilatation (or radial pulsation)
degree of freedom, while $\xi\neq 1$ corresponds to spheroidal
deformations and $\tau\neq 0$ to triaxial deformations.

The velocity of the particles are
\begin{eqnarray}
\label{eq.velocity.vectors.dep.2}
\vecv_i (\dot\lambda,\dot\xi, \dot\tau,\lambda,\xi,\tau) = R\dot
\lambda[ (\xi+\tau) x^*_i \vece_x +(\xi-\tau) y^*_i\vece_y
+(\xi^2- \tau^2)^{-1} z^*_i \vece_z] \\
+R \lambda[ (\dot \xi+\dot \tau) x^*_i \vece_x + (\dot \xi-\dot
\tau) y^*_i\vece_y -2(\xi\dot \xi -\tau\dot\tau )(\xi^2-
\tau^2)^{-2} z^*_i \vece_z ] .
\end{eqnarray}
An elementary computation shows that the kinetic energy,
$T=\frac{1}{2}\sum_{i } m_{i}\vecv_{i}^{2} $ , becomes
\begin{eqnarray}
\label{eq.kin.energy.2}
T(\dot\lambda,\dot\xi, \dot\tau,\lambda,\xi,\tau) =\frac{I}{2} \left[
\left(\xi^2+\tau^2 +\frac{1}{2\delta^2} \right) \dot\lambda^2 +
2\lambda\xi \left(1-\frac{1}{\delta^3}\right) \dot\lambda \dot \xi
+\right.  \\
\left.  2\lambda\tau\left(1+\frac{1}{\delta^3} \right) \dot\lambda
\dot \tau + \nonumber \lambda^2\left(1+\frac{2\xi^2}{\delta^4} \right)
\dot\xi^2 + \lambda^2\left(1+\frac{2\tau^2}{\delta^4} \right)
\dot\tau^2 - 4\lambda^2\frac{\xi\tau}{\delta^4} \dot\xi\dot\tau
\right]
\end{eqnarray}
Here $\delta \equiv \xi^2-\tau^2$, and $I=2R^2 \sum_{i}m_{i}x_{i}^{*2}
= MR^2 \kappa^2$.

The potential energy is given in Eq.\ (\ref{eq.energy.xi.tau}).
To get an explicit potential energy we must find some expression for
the volume dependent (pressure) energy $\Phi_{v}(R\lambda)$.  As a
model for this part of the energy we take the simple expression
\begin{equation}
\label{eq.vol.dep.energy}
\Phi_{v}= \frac{1}{n+1} \frac{GM^2}{\nu R} \frac{1}{\lambda^{n+1}}
\end{equation}
where we must take $n>0$ if the corresponding force is to balance
gravity and prevent collapse.  Putting these together we thus find the
total potential energy
\begin{equation}
\label{eq.tot.pot.energy.triaxial}
U(\lambda,\xi,\tau)= \frac{1}{n+1} \frac{GM^2}{\nu R}
\frac{1}{\lambda^{n+1}} +\frac{GM^2}{\nu R}\frac{1}{\lambda}
\chi(\xi,\tau) -\frac{1}{2} MR^2\kappa^2 \omega^2 \lambda^2
(\xi^2+\tau^2),
\end{equation}
If we use the definition of $k$ in Eq.\ (\ref{eq.def.k})
we find the expression
\begin{equation}
\label{eq.tot.pot.energy.triaxial.k}
U(\lambda,\xi,\tau)= \frac{GM^2}{\nu R} \left(
\frac{1}{(n+1)\lambda^{n+1}} +\frac{1}{\lambda} \chi(\xi,\tau)
-\frac{1}{2} k \lambda^2 (\xi^2+\tau^2) \right),
\end{equation}
for the potential energy of the system.

Combining $T$ of Eq.\ (\ref{eq.kin.energy.2}) with $U$ to the
Lagrangian $L(\dot\lambda,\dot\xi, \dot\tau,\lambda,\xi,\tau)=T-U$ we
can get the dynamics of this three degree of freedom system exactly by
finding and solving the Euler-Lagrange equations of the system.  Here
we will instead assume small oscillations and make a normal mode
analysis.  We first taylor expand $U$ to quadratic terms around
$\lambda=\xi=1, \tau=0,$ and then solve for the zeros of the first
derivatives of this quadratic to get the position of the minimum.  The
result is
\begin{eqnarray}
\label{min.pos,lambda}
\lambda_{0}&=&1+\frac{k(8+5k)}{8(n-k)-5(n+3k)k}\approx 1+\frac{k}{n},\\
\label{min.pos,xi}
\xi_{0}&=&1+\frac{5k(n+k)}{8(n-k)-5(n+3k)k}\approx 1+\frac{5}{8}k,\\
\tau_{0} &=&0 .
\end{eqnarray}
We now introduce new variables, $\rho, \sigma$ through
 \begin{equation}
\label{eq.small.amplitude.variables}
\rho=\lambda-\lambda_{0},\; \sigma=\xi-\xi_{0}; .
\end{equation}
Then, of course, $\dot\rho=\dot\lambda, \dot \sigma=\dot \xi$.  In the
kinetic energy we replace $\lambda, \xi, \tau,$ in the coefficients
with the equilibrium positions $\lambda_{0}, \xi_{0}, \tau_{0}$ and
expand to first order in $k$.  In the potential energy we make the
same replacement in the quadratic taylor expansion, expand to first
order in $k$ and throw away constant terms.  The result is that
 \begin{equation}
\label{eq.small.amplitude.T}
T=\frac{MR^2\kappa^2}{2} \left[ \frac{3}{2} \dot\rho^2+ 3\dot \sigma^2
+\dot \tau^2 + \frac{15k}{2}( \dot\rho\dot \sigma-\dot \sigma^2) +
\frac{2k}{n}(3\dot \sigma^2 +\dot \tau^2) \right]
\end{equation}
and that
 \begin{equation}
\label{eq.small.amplitude.U}
U=\frac{GM^2}{2R\nu} \frac{8}{15} \left[ \frac{15n}{8} \rho^2
+3\sigma^2+\tau^2 -\frac{15k}{8}(\rho^2 +4\rho\sigma +\sigma^2+\tau^2)
\right] .
\end{equation}
Only the degrees of freedom $\rho$ and $\sigma$ are coupled and only
when $k>0$.  For $k=0$ the spheroidal $\sigma$-mode and the triaxial
$\tau$-mode are degenerate (have the same frequency).  The value
$n=4/5$ is special since for that value all three modes have the same
frequency.

For the $\rho$-mode pressure is restoring force in the contracting
phase of the motion and gravity in the expanding phase.  for the other
two modes, $\sigma, \tau$, gravity alone acts to restore the spherical
minimum.  If we put
 \begin{equation}
\label{eq.Omega.def}
\Omega^2_{0}\equiv \frac{8GM}{15R^3\nu\kappa^2}
=\frac{GM}{R^3}(2-3\kappa^2)
\end{equation}
Eq.\ (\ref {eq.def.k}) shows that
$k=\frac{8}{15}\frac{\omega^2}{\Omega^2_{0} }$, and Eq.\
(\ref{eq.maclaurin.approx.eq}), with $e^2\approx 2\epsilon$, that
$k=\frac{8}{15}\epsilon$.  For $n>4/5$ we then get the approximate
eigen frequencies
\begin{eqnarray}
\omega^2_{\rho} =\frac{5n}{4}\Omega^2_{0} -\frac{2}{3}\omega^2,\\
\omega^2_{\sigma}
=\Omega^2_{0}+\left(1-\frac{16}{15n}\right)\omega^2,\\
\omega^2_{\tau} =\Omega^2_{0}-\left(1+\frac{16}{15n}\right)\omega^2,
\end{eqnarray}
to first order in $k$.  From this one easily finds the following first
order results
\begin{eqnarray}
T_{\rho}
=\frac{2\pi}{\Omega_{0}}\frac{2}{\sqrt{5n}}\left(1+\frac{4}{15n}\epsilon\right),\\
T_{\sigma} =\frac{2\pi}{\Omega_{0}}
\left[1-\frac{1}{2}\left(1-\frac{16}{15n}\right)\epsilon\right],\\
T_{\tau} =\frac{2\pi}{\Omega_{0}}
\left[1+\frac{1}{2}\left(1+\frac{16}{15n}\right)\epsilon\right],
\end{eqnarray}
for the corresponding periods.

The free radial oscillation mode for the Earth is known \cite{udias}
to have $T_{\rho}=20.45\,$min = $1228\,$s.  This means that one can
calculate $n$ and find it to be $n=13.19$ for the Earth.  Since the
equilibrium radius of the rotating Earth is $R\lambda_{0}$ Eq.\
(\ref{min.pos,lambda}), and a small calculation, now shows that
Earth's mean radius is $863$ m larger due to rotation compared to the
non-rotating case.  The fairly large $n$ also shows that the $g$ mode
periods are essentially independent of $n$.  For the Earth one finds
that the $\tau$-mode has period $T_{\tau}=83.22$ min.  Due to the
rotational splitting, which is given by
 \begin{equation}
\label{eq.rotational.splitting}
\Delta T=T_{\tau}-T_{\sigma}= {2\pi} \epsilon /{\Omega_{0}},
\end{equation}
the $\sigma$-mode is $17$ seconds shorter.  The longest observed free
oscillation period of the Earth has period $53.8\,$min (see Ud\'{\i}as
\cite{udias}).  The main explanation for the discrepancy is probably
that the model neglects elastic forces.  Though not quantitatively
reliable from this point of view the model has the advantages of
showing in a simple way how rotational splitting of the modes arise
and the order of magnitude of such splittings.

\section{Conclusions}
Some apparently new results relating to the classic theory of the
figure of rotating bodies have been presented.  The basic model, a
point mass at the center of a homogeneous fluid, is characterized by
their mass ratio, and interpolates between the limits of an
ellipsoidal homogeneous fluid and a body dominated by a small central
mass concentration.  It allows simple analytic treatment but is still
flexible enough to correctly describe the essential hydrostatics of
real rotating planets as well as stars.  Such models are always
useful, especially when one wishes to analyze, compare, and understand
large numbers of observational data.  In spite of the simplicity there
is no perturbation order to which the results are valid; the
nonlinearity of the basic equations can be retained.  To be more
precise Eq.\ (\ref{eq.maclaurin.approx}) shows that the formulas are
valid to seventh order in the eccentricity (within the basic model
with its simplified mass distribution).  As demonstrated by the
numerical experiments on Jupiter and Saturn data, this is essential.
In fact Table~\ref{table} indicates that the geometric oblateness of
the surface equipotential surface of Mars and Uranus determined from
observed $q$ and $J_{2}$ values using
(\ref{eq.maclaurin.approx.relation.epsilon}) probably are more
reliable than current observational data.  Finally the dynamics of the
model reveals the essentials of the coupling and rotational splitting
of the most basic free oscillations of a planet without recourse to
expansion in spherical harmonics.\\
\\
\noindent {\bf Acknowledgement} Constructive comments from Dr.\ John
D.\ Anderson on a previous version of this manuscript are gratefully
acknowledged.

\bibliographystyle{plain}

\newpage

\begin{table*}[t]

\begin{tabular}{|l|l|l|l|l|l|l|} \hline \hline
body  & $q$  & $\epsilon$  & $J_2$ & $e^2$ & $\kappa^2$
\\
\hline
{\protect\scriptsize Sun$^4$} &
{\protect\scriptsize $1.15 \!\cdot\! 10^{-5}$} &
{\protect\scriptsize $6.3 \!\cdot\! 10^{-6}$} &
{\protect\scriptsize $3.7 \!\cdot\! 10^{-7}$} &
{\protect\scriptsize $1.3 \!\cdot\! 10^{-5}$} &
{\protect\scriptsize $0.059$}
\\
\hline
{\protect\scriptsize Earth$^1$} &
{\protect\scriptsize $3.45 \!\cdot\! 10^{-3}$} &
{\protect\scriptsize $3.35 \!\cdot\! 10^{-3}$} &
{\protect\scriptsize $1.08 \!\cdot\! 10^{-3}$} &
{\protect\scriptsize $6.69 \!\cdot\! 10^{-3}$} &
{\protect\scriptsize $0.3307$}
\\
{\protect\scriptsize Earth$^2$} &
{\protect\scriptsize $3.45 \!\cdot\! 10^{-3}$} &
{\protect\scriptsize $3.35 \!\cdot\! 10^{-3}$} &
{\protect\scriptsize $1.08 \!\cdot\! 10^{-3}$} &
{\protect\scriptsize $6.69 \!\cdot\! 10^{-3}$} &
{\protect\scriptsize $0.323$}
\\ \hline
{\protect\scriptsize Mars$^1$} &
{\protect\scriptsize $4.57 \!\cdot\! 10^{-3}$ }  &
{\protect\scriptsize $6.48\!\cdot\! 10^{-3}$ } &
{\protect\scriptsize $1.96\!\cdot\! 10^{-3}$ }   &
{\protect\scriptsize $1.29 \!\cdot\! 10^{-2}$} &
{\protect\scriptsize $0.366$}
\\
{\protect\scriptsize Mars$^2$} &
{\protect\scriptsize $7.04 \!\cdot\! 10^{-3}$ }  &
{\protect\scriptsize $5.24\!\cdot\! 10^{-3}$ } &
{\protect\scriptsize $2.78\!\cdot\! 10^{-3}$ } &
{\protect\scriptsize $1.04\!\cdot\! 10^{-2}$ } &
{\protect\scriptsize $0.375$ }
\\ \hline
{\protect\scriptsize Jupiter$^1$} &
{\protect\scriptsize $8.34 \!\cdot\! 10^{-2}$ }  &
{\protect\scriptsize $6.49\!\cdot\! 10^{-2}$ } &
{\protect\scriptsize $1.47\!\cdot\! 10^{-2}$ } &
{\protect\scriptsize $1.26\!\cdot\! 10^{-1}$ } &
{\protect\scriptsize $0.254$ }
 \\
{\protect\scriptsize Jupiter$^2$} &
{\protect\scriptsize $8.29 \!\cdot\! 10^{-2}$ }  &
{\protect\scriptsize $6.51\!\cdot\! 10^{-2}$ } &
{\protect\scriptsize$1.45\!\cdot\! 10^{-2}$ }  &
{\protect\scriptsize $1.26\!\cdot\! 10^{-1}$ } &
{\protect\scriptsize $0.233$ }
\\
{\protect\scriptsize Jupiter$^3$ } &
{\protect\scriptsize $8.56 \!\cdot\! 10^{-2}$ }  &
{\protect\scriptsize $6.37\!\cdot\! 10^{-2}$ } &
{\protect\scriptsize $1.54\!\cdot\! 10^{-2}$ }  &
{\protect\scriptsize $1.27\!\cdot\! 10^{-1}$ } &
{\protect\scriptsize $ $ }
\\ \hline
{\protect\scriptsize Saturn$^1$} &
{\protect\scriptsize $1.40 \!\cdot\! 10^{-1}$ }  &
{\protect\scriptsize $9.80\!\cdot\! 10^{-2}$ } &
{\protect\scriptsize $1.63\!\cdot\! 10^{-2}$ }  &
{\protect\scriptsize $1.86\!\cdot\! 10^{-1}$ } &
{\protect\scriptsize $0.210$ }
\\
{\protect\scriptsize Saturn$^2$} &
{\protect\scriptsize $1.41 \!\cdot\! 10^{-1}$ }  &
{\protect\scriptsize $9.73\!\cdot\! 10^{-2}$ } &
{\protect\scriptsize $1.67\!\cdot\! 10^{-2}$ }   &
{\protect\scriptsize $1.85\!\cdot\! 10^{-1}$ } &
{\protect\scriptsize $0.176$ }
\\
{\protect\scriptsize Saturn$^3$ }&
{\protect\scriptsize $1.47 \!\cdot\! 10^{-1}$ }  &
{\protect\scriptsize $9.44\!\cdot\! 10^{-2}$ } &
{\protect\scriptsize $1.87\!\cdot\! 10^{-2}$  }  &
{\protect\scriptsize $1.89\!\cdot\! 10^{-1}$ }  &
{\protect\scriptsize $ $ }
\\ \hline
{\protect\scriptsize Uranus$^1$} &
{\protect\scriptsize $2.89 \!\cdot\! 10^{-2}$  } &
{\protect\scriptsize $2.29\!\cdot\! 10^{-2}$ }&
{\protect\scriptsize $3.52\!\cdot\! 10^{-3}$ } &
{\protect\scriptsize $4.53\!\cdot\! 10^{-2}$ } &
{\protect\scriptsize $0.225$ }
\\
{\protect\scriptsize Uranus$^2$} &
{\protect\scriptsize $3.50 \!\cdot\! 10^{-2}$ }  &
{\protect\scriptsize $1.98\!\cdot\! 10^{-2}$ }&
{\protect\scriptsize $5.55\!\cdot\! 10^{-3}$ }   &
{\protect\scriptsize $3.92\!\cdot\! 10^{-2}$ } &
{\protect\scriptsize $0.179$ }
\\ \hline
{\protect\scriptsize Neptune$^1$} &
{\protect\scriptsize $2.56 \!\cdot\! 10^{-2}$ }  &
{\protect\scriptsize $1.71\!\cdot\! 10^{-2}$ } &
{\protect\scriptsize $3.54\!\cdot\! 10^{-3}$ }  &
{\protect\scriptsize $3.39\!\cdot\! 10^{-2}$ } &
{\protect\scriptsize $0.24$ }
\\
{\protect\scriptsize Neptune$^2$} &
{\protect\scriptsize $2.34 \!\cdot\! 10^{-2}$ }  &
{\protect\scriptsize $1.82\!\cdot\! 10^{-2}$ } &
{\protect\scriptsize $2.80\!\cdot\! 10^{-3}$ }   &
{\protect\scriptsize $3.61\!\cdot\! 10^{-2}$ } &
{\protect\scriptsize $0.196$ }
\\
\hline
\end{tabular}

\caption{
Values of $q=R^3\omega^2/GM$, of oblateness,
$\epsilon$, and gravitational quadrupole, $J_{2}$,
ellipticity, $e$, squared  and dimensionless moment of inertia $\kappa^2=I/MR^2$.
}
\label{table}
\end{table*}

{\footnotesize
\noindent 1. The first row for each planet gives literature \protect\cite{lodders}
 data. These are  observational except $\kappa^2$ which are based on
theoretical calculations.
\protect\\
2. The second row for each planet gives the corresponding calculated
values as given by formulas (\protect\ref{eq.maclaurin.approx.relation.q}) -
(\protect\ref{eq.eps.e2}) in such a way that for each pair of observational $q$,
$J_{2}$ and $e^2$ the missing third is calculated. The moment of inertia is
calculated from observational $q$ and $J_{2}$ using formula
(\protect\ref{eq.J2.expression}), $\kappa^2=2 J_{2}/e^2$, with
$e^2(q,J_{2})$ from formula (\protect\ref{eq.maclaurin.approx.relation.epsilon}).
\protect\\
3. The third row for Jupiter and Saturn gives data calculated in
a similar way to that in the second row except that the first order
formula (\protect\ref{eq.J2.alpha.epsilon}) has been used to find the third
value from two observational values.
\protect\\
4. For the Sun $\epsilon$, $J_{2}$ and $e^2$ have been calculated
from an observation based $q$-value discussed in the text and a
theoretical $\kappa^2$  \protect\cite{lodders}, using formulas
(\protect\ref{eq.ellipt.kappa2}), (\protect\ref{eq.J2.kappa2}) and
$e^2=2\epsilon$ respectively.
}

\end{document}